\documentclass[]{revtex4}
\usepackage{hyperref}
\usepackage{color} 

\begin{document}

\title{  BRST symmetry and Darboux transformations in Abelian 2-form gauge theory }

 \author{ Sudhaker Upadhyay}
 \email {  sudhakerupadhyay@gmail.com; 
 sudhaker@boson.bose.res.in}

\affiliation { S. N. Bose National Centre for Basic Sciences,\\
Block JD, Sector III, Salt Lake, Kolkata -700098, India. }

\begin{abstract}
  We analyse the constraints of an Abelian 2-form gauge theory using Faddeev-Jackiw  symplectic formalism.
Further, this theory is treated  as a constrained system   in the context 
of Batalin-Fradkin-Vilkovisky  formalism to retrieve the BRST
symmetry. Using  the
fields decompositions the effective action
for Abelian 2-form gauge theory is written 
in terms of diagonalized uncanonical part and BRST exact one.
The nilpotent BRST and contracting homotopy $\sigma $ closed 
transformations with
field redefinitions are shown as the Darboux transformations used 
 in the Faddeev-Jackiw   formalism.
\end{abstract}

\maketitle 
\section{introduction}
The quantization of non-singular systems is in principle straightforward.
On the other hand, the quantization of singular systems (i.e. systems with constraints)
is non-trivial.
 The generalized Hamiltonian dynamics of singular 
systems  was initiated by Dirac \cite{di1, di2}. The dynamics of such systems are widely used in 
investigating theoretical models in contemporary elementary particle physics \cite{ht}.
Dirac proposed a kind of bracket to quantize these (singular) systems.
In the Dirac approach of dealing with singular systems, the dynamical equations
involve the variables of the entire phase space, including also unphysical gauge
degrees of freedom.  However,  a symplectic approach  to quantize the singular systems has been used by Faddeev and Jackiw (FJ)  \cite{b}, so-called
FJ approach,  
in which the systematic algorithm involves only the physical
(unconstrained) degrees of freedom for
arriving at a set of Hamilton equations of motion  \cite{rothe}.
In Ref. \cite{pons}, this algorithm has been  shown equivalent to Dirac approach.
In  FJ approach, the  Lagrangian is treated in (symplectic) first order.

Abelian antisymmetric rank-2 tensor field theory is an example of singular system 
where some of the constraints are not independent and said to be reducible.
This theory is the subject of interests in various aspects \cite{kara,kaul,sud,sud2, suga,rl,sase,grsc}. For example, 
  Kalb and Ramond has shown that   Abelian rank-2 antisymmetric fields interact 
with classical strings \cite{kara}, which was further applied to the dual description of 
Abelian Higgs model \cite{suga,rl}. The antisymmetric tensor field appears to couple the gravity or 
supergravity fields with higher curvature term in four and ten dimensions \cite{sase} and 
complete understanding of these couplings in superstring theories are crucial in order to 
have anomalies cancellation \cite{grsc}. 
Abelian 2-form gauge fields have their relevance in M-theory also. 
In particular,  the action for multiple M2-branes was studied via BLG
theory  \cite{bag,bag1,gus,fai, 1fai}. The gauge symmetry of this theory was generated by 
a Lie $3$-algebra rather than a Lie algebra. However, this limited the scope of this thoery 
to two M2-branes. 
So, this theory  was generalized to the ABJM theory \cite{ori, 
bu,nas, fai1,fai2}. The gauge symmetry of this theory is generated by the gauge 
 group $U(N) \times U(N)$.
The BRST symmetry of the ABJM theory has also been studied \cite{fai3, 1fai3, 2fai3}.
The ABJM theory has been generalized to the theory of
fractional M2-branes \cite{aha, klu}. The gauge group of ABJ theory is 
$U(N) \times U(M)$. 
Recently, the
BRST symmetry of the ABJ theory has also been studied \cite{fai4, 1fai4, 2fai4}.   
It is shown in \cite{pm} at quadratic order of the Lagrangian that the M5-brane
theory contains a self-dual two-form gauge field, in addition to the scalars 
corresponding to fluctuations of the M5-brane in the transverse directions, as well as
their fermionic super-partners. 
 The symplectic   quantization for   Abelian rank-2 antisymmetric
tensor field theory has been done in Ref. \cite{neto,neto1}. However, the 
Darboux transformations is not studied for the Abelian rank-2 antisymmetric field in the FJ context.  
This provides a motivation for this present work.

Batalin-Fradkin-Vilkovisky (BFV)  formulation  is a  Hamiltonian path integral approach
to quantize the constrained systems \cite{frvi,bv}. In this approach one extends 
the phase space   of the
theory   by introducing a conjugate momentum
for every Lagrange multiplier and  a ghost field for every constraint.
The induced effective action in extended phase space
exhibits a so-called BRST symmetry \cite{brst}. 
 However, in FJ approach the phase space is reduced by iteratively solving the 
constraints and performing the Darboux transformations, until we end up with an 
unconstrained and canonical Lagrangian.
The relation between BFV   quantization
scheme and the FJ approach for gauged SU(2) WZW model has been established in \cite{pa}.
 We explore it  for the reducible gauge theory  of Abelian rank-2 tensor field.

In this work we start with the FJ constraint analysis for 2-form gauge theory. The 
constraints which we
found (primary and zero iterated)  are exactly same as obtained from Dirac analysis but in more elegant manner. Then, we 
use BFV approach by extending the phase space
to analyse the BRST symmetry of the effective action. Further,  the BFV action is 
written 
in two terms, the first one is uncanonical term that we would obtain with the FJ 
method
after solving the constraints and the second one is BRST exact term.
The BRST transformation and contracting homotopy $\sigma $ closed transformations are calculated 
for the reducible 2-form gauge theory. Under the fields decompositions
these transformations are shown as the Darboux transformation used in FJ formalism. 

The  paper is organized as follows. In Sec II, we discuss the preliminaries   of the FJ 
symplectic approach of singular system. In Sec. III, we make an analysis to
investigate the constraints structure of Abelian  rank-2 tensor field theory
using symplectic matrix. Then,  we stress the BRST-BFV 
formulation to quantize such reducible gauge theory  in sec. IV. Further, in Sec. V,
we show that the BRST and contracting homotopy $\sigma$ transformations of 2-form gauge theory are basically Darboux transformations used in FJ symplectic approach. The last section is kept for 
making  concluding remarks.
\section{ Faddeev-Jackiw approach: general formulation}
In this section we discuss the methodology of FJ approach to quantize the singular
systems.
In this formalism, we first write 
the Lagrangian of a singular system   into the
first-order form as follows:
\begin{equation}
L(\xi )=a_i(\xi )\dot{\xi}^i -V(\xi ), \ \ (i=1, 2, 3, .....,n), \label{lag}
\end{equation}
where $\xi^i$ is called the symplectic variable, $V (\xi )$ is called the symplectic potential.
The first-order form 
can be implemented by introducing some auxiliary variables
($a_i$) such as the canonical
momentum \cite{mon}. The Euler-Lagrange equations of motion for Lagrangian (\ref{lag}) can be written as
\begin{equation}
f_{ij}(\xi)\dot{\xi}^j=\frac {\partial V (\xi )}{\partial \xi^i}\ \ \ (i=1, 2, 3,.....,n),
\label{eom1}
\end{equation}
where $f_{ij}$ is  so-called  symplectic matrix
with following explicit form:
\begin{equation}
f_{ij}(\xi )=\frac{\partial a_j}{\partial \xi^i}-\frac{\partial a_i}{\partial \xi^j}.
\end{equation}
If matrix $f_{ij}$ is regular (invertible), all symplectic variables can 
be solved
from (\ref{eom1})
\begin{equation}
\dot{\xi}^j=f_{ij}^{-1}\frac {\partial V (\xi )}{\partial \xi^i}\ \ \ (i=1, 2, 3,.....,n).
\end{equation}
If matrix $f_{ij}$ is singular, there are some constraints in this system.
In order to quantize the
system with constraints in the FJ method, Barcelos-Neto and Wotzasek \cite{bnw, bnw1}
 proposed the symplectic algorithm to extend the original FJ method \cite{b}.
We give a brief description of the symplectic algorithm here. The constraints arising from
Eq. (\ref{eom1}) are
\begin{equation}
\Omega_\alpha^{(0)}= (U_\alpha )_i\frac{\partial V}{\partial \xi^i}=0 \ \ \ 
(\alpha =1, 2, 3,....,m),
\end{equation}
where $U_\alpha$ is the zero mode of the symplectic matrix $f$, $m = n - r $ ($r$ is the rank of $f$)
\begin{equation}
(U_\alpha)^T f=0,
 \ \ (\alpha =1, 2, 3,....,m).
\end{equation}
Now, we modify our original Lagrangian by introducing the constraint term multiplied with 
some Lagrange multipliers ($v^\alpha$) as
\begin{equation}
L_{mod} =a_i(\xi )\dot{\xi}^i -V(\xi )
-v^\alpha \Omega_\alpha^{(0)}, \ \ (\alpha =1, 2, 3,....,m)
\end{equation} 
and calculate the symplectic matrix with modified Lagrangian density; 
if there is further constraint in the system then the matrix be-
comes singular otherwise it is nonsingular.
But, doing iteratively, in the last we get a nonsingular matrix. This means there is no further 
constraint in the system. So
according to the Darboux theorem \cite{db} there exists
a coordinate transformation 
\begin{eqnarray}
Q_1(\xi^{(0)}),......,Q_{m/2}(\xi^{(0)});\\
P_1(\xi^{(0)}),......,P_{m/2}(\xi^{(0)}),
\end{eqnarray}
which transform the first-order Lagrangian given in Eq. (\ref{lag}) into
\begin{equation}
L^{(0)}=P_k\dot{Q}_k-V^{(0)}(P,Q), \ \ (k=1, 2, ......m/2).
\end{equation}
From the
mathematical view, the key of the FJ method
is just to construct such a Lagrangian that satisfies the
Darboux theorem, and the FJ canonical
quantization is established on such a form of the
Lagrangian.

In the next section we will treat the  Abelian 2-form gauge theory as
a singular system and will analyse 
 the constraints using the FJ symplectic approach.
\section{Constraints analysis of Abelian 2-form gauge theory: using FJ approach}
We start with the Lagrangian density for Abelian free  Kalb-Ramond theory in (1+3) dimensions (4D) \cite{kara} given by
\begin{equation}
{\cal L}=\frac{1}{12} F_{\mu \nu \rho}F^{\mu \nu \rho},\label{kin}
\end{equation}
where the antisymmetric field strength tensor in terms of Kalb-Ramond field ($B_{\mu\nu}$) is defined as $F_{\mu\nu\lambda}=\partial_\mu 
B_{\nu\lambda}+ \partial_\nu B_{\lambda\mu}+ \partial_\lambda B_{\mu\nu}$.
This Lagrangian density is invariant under the
following gauge transformation 
\begin{equation}
\delta B_{\mu\nu} = \partial_\mu\Lambda_\nu -\partial_\nu\Lambda_\mu,
\end{equation}
where $\Lambda_\mu$ is an arbitrary vector parameter. 

This gauge transformation is reducible, since particular choice of vector parameter, i.e.
\begin{equation}
\Lambda_\mu = \partial_\mu \varepsilon,
\end{equation}
 leads to $\delta B_{\mu\nu} =0$.

Now,   the canonical momenta corresponding to the fields $B_{0i}$ and $B_{ij}$, respectively, are calculated as
\begin{equation}
\Pi^{0i}=\frac{\partial{\cal L}}{\partial\dot B_{0i}}=0,
\end{equation}
and
\begin{equation}
\Pi^{ij}=\frac{\partial{\cal L}}{\partial\dot B_{ij}}=\frac{1}{2}F^{0ij}.
\end{equation}
The primary constraint of the theory  thus obtained is
\begin{equation}
\Pi^{0i}\approx 0.
\end{equation}
Now, in order to make  the Lagrangian density for Abelian 2-form gauge theory in the 
first order symplectic form, we calculate the following 
\begin{equation}
\Pi^{\mu\nu}\dot B_{\mu\nu} -{\cal L}=\Pi^{ij}\Pi_{ij}+\frac{1}{12} F_{ijk}F^{ijk}
+2\Pi^{ij}\nabla_i B_{0j}.
\end{equation}
So, the first order symplectic version of Lagrangian density given in 
Eq. (\ref{kin}) is given by
\begin{eqnarray}
{\cal L}^{(0)}&=&\Pi^{\mu\nu}\dot B_{\mu\nu}-\Pi^{ij}\Pi_{ij}-\frac{1}{12} F_{ijk}F^{ijk}
-2\Pi^{ij}\nabla_i B_{0j},\nonumber\\
& =&\Pi^{ij}\dot B_{ij}-V^{(0)},
\end{eqnarray}
where $V^{(0)}$ is the symplectic potential with following expression:
\begin{equation}
V^{(0)}=\Pi^{ij}\Pi_{ij}+\frac{1}{12} F_{ijk}F^{ijk}
+2\Pi^{ij}\nabla_i B_{0j}.
\end{equation}
The corresponding 
symplectic equations of motion can be calculated easily with
following relations
\begin{equation}
f_{ijk\lambda}^{(0)}\dot\xi^{k\lambda}=\frac{\partial V^{(0)}(\xi )}{\partial \xi^{ij}},
\end{equation}
where 
\begin{equation}
f_{ijk\lambda}^{(0)}=\frac{\partial a_{k\lambda}({\bf y})}{\partial \xi^{ij}({\bf x})}-
\frac{\partial a_{ij}({\bf x})}{\partial \xi^{k\lambda}({\bf y})}.
\end{equation}
The set of symplectic variables are
\begin{equation}
\xi^{(0)}(x)=\{B_{ij}, \Pi_{ij}, B_{oi}\}.
\end{equation}
The components of symplectic 1-form are calculated as follows:
\begin{eqnarray}
a_{B_{ij}}^{(0)}&=&\frac{\partial{\cal L}}{\partial\dot B_{ij}}=\Pi^{ij},\nonumber\\
a_{\Pi_{ij}}^{(0)}&=&\frac{\partial{\cal L}}{\partial\dot \Pi_{ij}}=0,\nonumber\\
a_{B_{0i}}^{(0)}&=&\frac{\partial{\cal L}}{\partial\dot B_{0i}}=0.
\end{eqnarray}
Thus the matrix $f^{(0)}$, whose general form reads
\begin{equation}
f^{(0)}=\left (\begin{array}{clcr}
         f_{ik}^{(0){B_{0i}{B_{0k}}}}&f_{ikl}^{(0){B_{0i}{B_{kl}}}}&f_{ikl}^{(0){B_{0i}{\Pi_
                                                           {kl}}}}\\
         f_{ijk}^{(0){B_{ij}{B_{0k}}}} &f_{ijkl}^{(0){B_{ij}{B_{kl}}}}&f_{ijkl}^{(0){B_{ij}{
\Pi_{kl}}}}\\
         f_{ijk}^{(0){\Pi_{ij}{B_{0k}}}}&f_{ijkl}^{(0){\Pi_{ij}{B_{kl}}}}&f_{ijkl}^{(0){\Pi_
{ij}{\Pi_{kl}}}}
         \end{array}\right),
\end{equation}
is calculated in this case as
\begin{equation}
f^{(0)}=\left (\begin{array}{clcr}
         0& \ \ \ \ \ \ \ \ 0& 0\\
         0&  \ \ \ \ \ \ \ \   0& \frac{1}{2}(-\delta_{ik}\delta_{jl}+\delta_{il}\delta_{jk})\\
        0&  \frac{1}{2}(\delta_{ik}\delta_{jl}-\delta_{il}\delta_{jk})& 0
         \end{array}\right)\delta^3 ({
\bf x}-{\bf y}).
\end{equation}
It is a singular matrix. The zero mode of this matrix is, 
 $(0, 0, \nu^{B_{0i}})$ where $\nu^{B_{0i}}$ is some arbitrary function. In terms of FJ method 
\cite{wot}, using zero-mode, we can obtain constraint
\begin{eqnarray}
\Omega^{i(0)}&=&(\nu^0)^T_{0i}\frac{\partial V^{(0)}}{\partial \xi^{0i}}\approx0,\nonumber\\
&=& \nu^{B_{0i}}\frac{\partial V^{(0)}}{\partial B^{0i}}\approx 0,\nonumber\\
&=& \nabla_j\Pi^{ij}\approx 0,
\end{eqnarray}
which is zero iterated constraint. This is not independent constraint, since it satisfies the reducibility  
condition $\nabla_i\Omega^{i(0)} =0 $.
However, it is easy to see that even after calculating the symplectic matrix
for modified Lagrangian density with above constraint
 the zero modes do not lead to any new constraint.
Hence, there is no further constraints in the theory.

We end  this section by concluding that both the constraints primary and zero iterated   are
exactly same as obtained in Dirac procedure.
\section{The extended action and BRST symmetry}
In the above section, we obtained two constraints (primary and zero iterated) in the 2-form gauge theory, i.e. $\Pi^{0i} =0$
and $\nabla_j\Pi^{ij}=0$. In this section we discuss the 
nilpotent BRST symmetry for Abelian rank-2 tensor field theory. To do so, we introduce
two pairs of canonically conjugate anticommuting ghosts $({\cal{C}}_i,{\cal{P}}_i)$ and $(\bar{{\cal{C}}}_i,\bar{{\cal{P}}}_i)$ corresponding to the above constraints. Further, we need the following  pairs of canonically conjugate commuting 
ghosts $(\beta, \Pi_\beta)$ and $({\bar{\beta}},\Pi_{\bar{\beta}})$ which are ghosts of ghosts according to  the property of reducibility. The  ghosts numbers of these ghost fields are as follows:
\begin{eqnarray}
gh ({\cal{C}}_i)&=&-gh({\cal{P}}_i)=1,\nonumber\\
 gh (\bar{\cal{C}}_i)&=&-gh (\bar{\cal
{P}}_i)=-1,\nonumber\\
gh (\beta )&=&-gh (\Pi_\beta ) =2,\nonumber\\
 gh(\bar{\beta})&=&-gh (\Pi_{\bar{\beta}})=-2,
\end{eqnarray}
and they satisfy the following (anti-)commutation relations
\begin{eqnarray}
\left\{{\cal{C}}_i({\bf x}),{\cal{P}}_j({\bf y}) \right\}&=&\; -i\; \delta_{ij}\;\delta^3 ({
\bf x}-{\bf y}),\nonumber\\
  \left\{\bar{{\cal{C}}}_i({\bf x}),{\bar{\cal{P}}}_j({\bf y}) 
\right\}&=&
\; -i\; \delta_{ij}\delta^3 ({\bf x}-{\bf y}),  \label{acom1}\\
\left[ \beta({\bf x}),\; \Pi_\beta ({\bf y})\right ]&=&\; i\delta^3 (\bf{x}-\bf{y}),\nonumber\\
 \left[ \bar{\beta}({\bf x}),\;\Pi_{\bar{\beta}} ({\bf y})\right] &=&\; i\;\delta^3 ({\bf 
x}-{\bf y}).\label{com3}
\end{eqnarray}
The phase space is further extended by introducing canonical conjugate pairs 
$({\cal{C}}_0,{\cal{P}}_0)$ and (${\bar{\cal{C}}}_0,\bar{\cal{P}}_0)$ as Lagrange multipliers 
to the pair $({\cal{C}}_i,{\cal{P}}_i)$, $(\bar{{\cal{C}}}_i,\bar{{\cal{P}}}_i)$ and a 
canonical pair ($\varphi,\Pi_{\varphi})$ as Lagrange multiplier to the gauge condition.
 Hence, the extended action is given by,
\begin{eqnarray}
S_{eff}&=&\int d^4x\left[\Pi^{0i}\dot B_{0i}+ \Pi^{ij}\dot B_{ij}+{\cal P}^i\dot {\cal C}_i 
+ \bar {\cal P}^i \dot {\bar{\cal C}}_i \right.\nonumber\\
&+&{\cal P}^0\dot {\cal C}_0 
+\left. \bar {\cal P}^0\dot {\bar{\cal C}}_0+\Pi_{\beta}\dot\beta +\Pi_{\bar\beta}\dot{
\bar\beta}
+\Pi_{\varphi}\dot\varphi -{ \cal H}_c -\{Q,\Psi\} \right],\label{effs}
\end{eqnarray}
where 
$\Psi $ is the gauge fixed fermion and $Q$ is the generator of the BRST symmetry. 
The canonical Hamiltonian density, ${\cal H}_c$, is calculated  as
\begin{equation}
{\cal H}_c=\Pi_{ij}\Pi^{ij} +\frac{1}{12}F_{ijk}F^{ijk}.
\end{equation}
The expression for BRST charge for Abelian 2-form gauge theory  is given by
\begin{eqnarray}
Q= -2\nabla_i\Pi^{ij}{\cal C}_j+\Pi_\varphi\bar{\cal P}_0-{\cal P}_0\Pi_{\bar\beta}
-\bar{\cal P}^i\Pi_{0i},\label{brsc}
\end{eqnarray}
which satisfies following algebra
\begin{eqnarray}
\{Q,Q\}&=&0, \ \ \{ {\cal H}_c, Q\} =0.
\end{eqnarray}
The ghost numbers of $Q$ and $\Psi $ are as follows:
\begin{eqnarray}
gh(Q)&=&1,\ \ \ gh(\Psi )=-1.
\end{eqnarray}
The BRST symmetry transformation can be calculated 
with following relation
\begin{equation}
s_b \phi=-i{\left[\phi,Q\right]}_\pm,
\end{equation} 
 where $+$ is used  for fermionic and $-$ for the
bosonic nature  of generic fields $\phi$ .
 Using the above relation and the expression for BRST charge given in Eq. (\ref{brsc}),
we calculated 
the BRST symmetry transformations for fields as follows:
\begin{eqnarray}
&&s_b  B_{ij}= \left(\nabla_i {\cal
{C}}_j-\nabla_j{\cal{C}}_i\right),\ \ s_b  B_{0i}=-{\bar{\cal{P}}}_i, 
 \nonumber\\
&&s_b  \Pi_{\varphi}=0,\ s_b {\cal{C}}_i =0, \ s_b {\bar{\cal
{C}}}_i=\Pi_{0i},\   s_b {\cal{C}}_0 =\Pi_{\bar{\beta}},
\nonumber\\
&&  s_b {\bar{\cal{C}}}_0 =\Pi_\varphi,\  \ s_b \varphi =-{\bar{\cal{P}}}_0,
\ s_b \beta =0,\nonumber\\
&&s_b \bar{\beta} =-{\cal{P}}_0,\ \ s_b \Pi_{0i}=0,
 \ \ s_b \Pi_{ij}=0,\nonumber\\
&&s_b {\cal{P}}^i  =2\nabla_j\Pi^{ji},\ \ s_b {\bar{\cal{P}}}_i
 =0,
 \ \ s_b {
\cal{P}}_0=0,\nonumber\\
&&s_b {\bar{\cal{P}}}_0=0, \ \ s_b \Pi_\beta =0,
\ \
s_b \Pi_{\bar{\beta}}=0.\label{qbrst}
\end{eqnarray}
These transformations are nilpotent (i.e. $s_b^2 =0$) and symmetry of the effective action  given in Eq. (\ref{effs}).
\section{BRST symmetry transformation as a Darboux transformation}
In this section, we study the BRST transformation and the contracting homotopy $\sigma$ transformation
for Abelian 2-form gauge theory under the Darboux transformation. For this purpose
we first decompose the field $B_{ij}$ into transverse and longitudinal parts as follows
\begin{eqnarray}
{ B}_{ij}&=&{B}_{ij}^T+ {B}_{ij}^L,\nonumber\\
&=&\epsilon_{ijk} \nabla_k B^T+ \nabla_i B_j^L -\nabla_j B^L_i, \label{b}
\end{eqnarray}
where ${B}_{ij}^T =\epsilon_{ijk} \nabla_k B^T$ and ${B}_{ij}^L =\nabla_i B_j^L -\nabla_j B^L_i$.
Then we decompose corresponding 
momenta $\Pi_{ij}$ into transverse and longitudinal parts as follows
\begin{eqnarray}
{\Pi}_{ij}&=&{ \Pi}_{ij}^T+{\Pi}_{ij}^L,\nonumber\\
&=&\epsilon_{ijk} \frac{\nabla_k}{\nabla^2
}{\Pi}^T+ \frac{1}{\nabla^2} \left[ \nabla_i\Pi_j^L -\nabla_j \Pi_i^L\right],
\label{pi}
\end{eqnarray}
where ${ \Pi}_{ij}^T = \epsilon_{ijk} \frac{\nabla_k}{\nabla^2
}{\Pi}^T$ and ${\Pi}_{ij}^L = \frac{1}{\nabla^2} \left[ \nabla_i\Pi_j^L -\nabla_j \Pi_i^L \right] $.
Further, we exploit the relations (\ref{qbrst}) to solve the
field variables ${\cal C}_i, \bar {\cal P}_i, \Pi^L_{ij}$ and
$ \Pi_{0i}$ in terms of BRST transformation as follows
\begin{eqnarray}
{\cal C}_i&=& s_b  B_i^L,\nonumber\\
\bar{\cal P}_i&=&s_b  B_{0i},\nonumber\\
\Pi^L_{ij}&=& \frac{\nabla_j}{2\nabla^2}s_b  {\cal P}_i,\nonumber
\\
\Pi_{0i}&=&s_b \bar{\cal C}_i.
\end{eqnarray}
Using the  fields decompositions the effective action given in Eq. (\ref{effs}) is
written as
\begin{eqnarray}
S_{eff}&=&\int d^4x \left[\Pi_{0i}\dot B^{0i}+\Pi_{ij}^T\dot B^{ijT}+ 2\frac{\nabla_i}{ 
\nabla^2 }\Pi_j^L \nabla^i \dot B^{jL}\right.\nonumber\\
&- &\left. 2\frac{\nabla_i}{ \nabla^2 }\Pi_j^L \nabla^j \dot B^{iL} +\dot {\cal C}_i {\cal P}^i+  \dot {\bar{\cal C}}_i\bar {\cal P}^i \right. \nonumber\\
&+&\left.\dot {\cal C}_0 {\cal P}^0 
+\dot {\bar{\cal C}}_0\bar {\cal P}^0+\Pi_{\beta}\dot\beta +\Pi_{\bar\beta}\dot{
\bar\beta} \right. \nonumber\\
&+&\left.\Pi_{\varphi}\dot\varphi -{\cal H}_c-\{ Q,\Psi\}\right ],
\end{eqnarray}
where the decomposed canonical Hamiltonian density is given by
\begin{eqnarray}
{\cal H}_c&=& \Pi^T_{ij}\Pi^{ijT}+ 2\frac{\nabla_i}{ 
\nabla^2 }\Pi_j^L \frac{\nabla^i}{ 
\nabla^2 } \Pi^{jL} \nonumber\\
&- & 2\frac{\nabla_i}{ \nabla^2 }\Pi_j^L \frac{\nabla^j}{ 
\nabla^2 } \Pi^{iL} +\frac{1}{12}F_{ijk}F^{ijk}.
\end{eqnarray} 
We can easily see that using the symmetry transformations 
the effective action for Abelian 2-form gauge theory 
can be recast as 
\begin{eqnarray} 
S_{eff}&= &\int d^4x\left[\Pi_{ij}^T\dot B^{ijT}+\Pi_\beta\dot\beta -{\cal H}+s_b \left(\bar{\cal C}^i\dot B_{0i}-{\cal P}^i \dot B_i^L \right.\right.\nonumber\\
&+&  \left. \left. {\cal C}_0\dot{\bar\beta}+
\bar{\cal C}_0\dot\varphi
+\frac{1}{4}s_b {\cal P}_i\frac{1}{\nabla^2}{\cal P}^i\right) - \{ Q,\Psi\}\right],
\end{eqnarray}
where 
\begin{equation}
{\cal H} = \Pi^T_{ij}\Pi^{ijT} +\frac{1}{12}F_{ijk}F^{ijk}.
\end{equation}

Hence, we can make the following choice for the gauge fermion 
\begin{equation}
\Psi = i\left(\bar{\cal C}^i\dot B_{0i}-{\cal P}^i \dot B_i^L 
+{\cal C}_0\dot{\bar\beta}+
\bar{\cal C}_0\dot\varphi +\frac{1}{4}s_b {\cal P}_i\frac{1}{\nabla^2}{\cal P}^i\right).
\end{equation}
Exploiting the canonical fields decompositions  given in Eqs. (\ref{b}) and (\ref{pi}), the nilpotent BRST symmetry transformations 
of Eq.  (\ref{qbrst}) have the following form:
\begin{eqnarray}
&&s_b B^L_i=  {\cal
{C}}_i,\ \ s_b B_{0i}={\bar{\cal{P}}}_i,\ \ s_b \Pi_{\varphi}=0, \nonumber\\
&&s_b{\cal{C}}_i =0, \ \ 
s_b{\bar{\cal{C}}}_i =\Pi_
{0i}, \ \ \
s_b{\cal{C}}_0 =\Pi_{\bar{\beta}},\nonumber\\
&&s_b{\bar{\cal{C}}}_0 =\Pi_\varphi,\  \ \ s_b\varphi =-{\bar{\cal{P}}}_0,
\ \ \
s_b\beta =0,\nonumber\\
&&s_b\bar{\beta} =-{\cal{P}}_0,  \ \ 
s_b\Pi_{0i}=0,  \ s_b\Pi^L_i =0, \nonumber\\
&& s_b{
\cal{P}}_0 =0,\ \ s_b{\cal{P}}_i  =2 \nabla^j \Pi_{ji}^L,\ \ s_b{\bar{\cal{P}}}_i =0,
\nonumber\\
&&s_b\Pi_\beta = 0,\ \ \ 
s_b{\bar{\cal{P}}}_0=0,
 \ \ \
s_b\Pi_{\bar{\beta}} =0,\nonumber\\
&&s_bB^T_{ij}=0,\ \ \ s_b \Pi^T_{ij}=0.
\end{eqnarray}
Here we notice that only transverse fields are BRST closed without being BRST exact. Therefore 
one can show that the functionals of these  transverse fields  are being used only in classical
BRST cohomology. 
The contracting homotopy $\sigma$ with respect to above BRST operator $s_b$ is
defined as
\begin{eqnarray}
&&\sigma( {\cal
{C}}_i)=  B^L_i,\  \ \sigma( B^L_i)=0, \ \ \sigma({\bar{\cal{P}}}_i)=B_{0i},  \nonumber\\
&& 
\sigma (B_{0i})=0,\ \ \sigma (\Pi_{0i})=\bar{\cal C}_i, \ \ \sigma (\bar{\cal C}_i)=0, \nonumber\\
&&\sigma (\Pi_{\bar\beta})={\cal C}_0,\ \sigma ({\cal C}_0)
=0, \ \ \sigma (\Pi_\varphi ) =\bar{\cal C}_0,
 \nonumber\\
&& \sigma (\bar{\cal C}_0)=0,\ \ \sigma (-\bar{\cal P}_0)=\varphi,\ \ \sigma (\varphi )=0,\nonumber\\
&& \sigma (-\ {\cal P}_0)=\bar{\beta},\ \ \sigma (\bar \beta)=0,\ \  \sigma (\beta)=0,\nonumber\\
&&\sigma\left(2 \nabla^j \Pi_{ji}^L\right)={\cal P}_i, \ \ \sigma ({\cal P}_i)=0,\nonumber\\
&& \sigma (\Pi_\beta )=0,\ \ \sigma (B^T_{ij})=0,\ \ \ \sigma (\Pi^T_{ij})=0,
\end{eqnarray}
which is also nilpotent in nature.
Further, $\sigma$ operator satisfies the following relation:
$\sigma s_b +s_b\sigma =N$, where $N$ counts the
degree in unphysical variables $B^L_i, {\cal {C}}_i, {\bar{\cal{P}}}_i, B_{0i},
 {\Pi_{0i}, \bar{\cal C}}_i, \Pi_{\bar\beta}, {\cal C}_0, \Pi_{\varphi},   \bar{\cal C}_0, \bar{\cal P}_0, \varphi, {\cal P}_0, \bar{\beta},
  \Pi_{ji}^L, {\cal P}_i,$ i.e.
\begin{eqnarray}
N&=&B^L_i \frac{\partial}{\partial B^L_i}+{\bar{\cal{P}}}_i\frac{\partial}{\partial {\bar{\cal{P}}}_i}+{\cal {C}}_i \frac{\partial}{\partial {\cal {C}}_i}+B_{0i} \frac{\partial}{\partial B_{0i}}+  \Pi_{0i} \frac{\partial}{\partial  {\Pi_{0i}}}
+\bar{\cal C} _i\frac{\partial}{\partial \bar{\cal C} _i}+\Pi_{\bar\beta} \frac{\partial}{\partial \Pi_{\bar\beta}}+{\cal C}_0 \frac{\partial}{\partial {\cal C}_0}\nonumber\\
&+& \Pi_{\varphi}\frac{\partial}{\partial  \Pi_{\varphi}}+\bar{\cal C}_0 \frac{\partial}{\partial \bar{\cal C}_0}+\varphi\frac{\partial}{\partial \varphi}+{\cal P}_0 \frac{\partial}{\partial {\cal P}_0}+ \bar{\beta} \frac{\partial}{\partial \bar{\beta}}+   \Pi_{ji}^L \frac{\partial}{\partial   \Pi_{ji}^L}+ {\cal P}_i \frac{\partial}{\partial  {\cal P}_i}.
\end{eqnarray} 
It follows that if the functional ${\cal G}$   of degree $n\neq 0$ is BRST closed 
in the unphysical variables,
\begin{eqnarray}
s_b{\cal G}=0,\ \ N{\cal G} =n {\cal G},
\end{eqnarray}
then it is BRST exact also, i.e. ${\cal G}=s_b[(1/n)\sigma {\cal G}]$.
However, only those BRST closed functionals,  which are of degree $n=0$ in the unphysical 
variables, are not BRST exact, i.e. the functionals of $B^T_{ij}, \Pi^T_{ij},
\beta, \Pi_\beta$ fields.

Therefore, the above BRST and $\sigma$  closed transformations
under which the fields transform are basically Darboux transformations
used in FJ quantization.
\section{conclusion}
We have considered the Abelian rank-2 antisymmetric tensor field theory
(which is a reducible gauge theory)
as a singular   system and  have investigated the constraints involved
 in the theory using the  FJ  symplectic approach.
  Further, we have implemented the  BFV 
formalism  in which the scalar potential, $B_{0i}$, is treated as a full dynamical variable with 
vanishing conjugate momentum, $\Pi_{0i}$. According to BFV formulation the phase space has been extended by introducing  
a canonical pair of ghost  fields for each constraint in the theory. The conserved BRST charge as well as BRST symmetry 
have been constructed for Abelian 2-form gauge theory
within Hamiltonian framework. We have shown that
using fields decompositions
the effective action for Abelian rank-2 tensor field theory can be written as a sum of an uncanonical 
term and the BRST exact one. Further, it has been shown that the field redefinitions under  which   the fields transform
into nilpotent  BRST and $\sigma$ closed transformations are basically Darboux transformations used in FJ approach.

  Further  use of similar analysis in the quantum theory
of gravity \cite{fd, mir1, 1mir1, 1mir, kon, es} and in higher derivative field theory \cite{moz} will be interesting. 
 It is also important to mention that  within the FJ framework the attempts to derive a
non-abelian version of this theory \cite{su} will be exotic.  

The path integral corresponding to the FJ quantization method has also been extensively studied under various aspects 
\cite{lh}. So far we have studied the  Darboux  transformations which appears in the FJ quantization  as a symmetry of such path integral. However, it will be interesting to explore the  Darboux  transformations under which   the 
path integral corresponding to the FJ quantization method is not invariant \cite{sud1}.

\end{document}